\documentclass[aps,pre,twocolumn,floatfix,showpacs]{revtex4}
\usepackage{epsf,graphicx,amsmath,amssymb}
\begin{document}
\title{Generalized priority-queue network dynamics: Impact of team and hierarchy}
\author{Won-kuk~Cho}
\affiliation{Department of Physics, Korea University, Seoul 136-713, Korea}
\author{Byungjoon~Min}
\affiliation{Department of Physics, Korea University, Seoul 136-713, Korea}
\author{K.-I.~Goh}
\thanks{Email: kgoh@korea.ac.kr}
\affiliation{Department of Physics, Korea University, Seoul 136-713, Korea}
\author{I.-M.~Kim}
\affiliation{Department of Physics, Korea University, Seoul 136-713, Korea}
\date{\today}
\begin{abstract}
We study the effect of team and hierarchy
on the waiting-time dynamics of priority-queue networks. 
To this end, we introduce generalized priority-queue 
network models incorporating interaction rules based on team-execution 
and hierarchy in decision making, respectively.
It is numerically found that the waiting time distribution exhibits 
a power law for long waiting times in both cases,
yet with different exponents depending on the team size and
the position of queue nodes in the hierarchy, respectively.
The observed power-law behaviors have in many cases a corresponding
single or pairwise-interacting queue dynamics,
suggesting that the pairwise interaction
may constitute a major dynamic consequence in the priority-queue networks.
It is also found that the reciprocity of influence 
is a relevant factor for the priority-queue network dynamics.
\end{abstract}
\maketitle

\section*{Introduction}
Priority-based queueing models are of interest
among statistical physics community recently 
\cite{alb,vazquez,caldarelli,grinstein,masuda,ov,min,anteneodo}.
They were originally introduced in the context
of human dynamics \cite{alb}, aiming to model individual's internal decision 
making process underlying observed human activity patterns,
yet bear implications to other problems in statistical physics
such as the Bak-Sneppen model 
for biological evolution \cite{bs} and invasion percolation \cite{caldarelli}
as well, from the extremal dynamics perspective. Thus the understanding 
of dynamics of the priority-queue models is of broader interest
beyond the human dynamics.

The primary fingerprint of priority-queue dynamics
is the heavy tail in the waiting time distribution $P(\tau)$, 
often taking an asymptotic power-law form for large waiting times,
\begin{equation} P(\tau)\sim \tau^{-\alpha},  \end{equation}
where the waiting time $\tau$ is the time interval between two consecutive events.
In the original single fixed-length queue model by Barab\'asi \cite{alb},
the exponent $\alpha\approx1$ was obtained numerically, which was
verified analytically later \cite{vazquez}. 
Such heavy-tailed dynamics have been observed for a range of
human activities such as the e-mail, library loan,
web browsing, and call initiation from a mobile phone \cite{alb,pre,goh-alb,phone2,amaral},
prompting further interests and debates.

Later on, the Barab\'asi model has been modified to include
different factors such as memory and multi-task executions
that would be of potential relevance in human decision
making \cite{memory,blanchard,web,activity}.
Among them, one of the most confounding factor would be the human 
interaction \cite{ov,min}: In a modern society, the activity of an individual
is rarely an outcome of completely autonomous decisions, but 
of delicate compromises and balanced conflicts between often competing
priorities of various settings in complex social networks \cite{rmp,guido-book,social}.
Beyond its sociological context, the human interaction introduces
coupling of multiple queue dynamics, rendering the model highly nontrivial.
Thus the study of the role of human interaction on priority-queue
dynamics is of theoretical interest as well.

There are a few studies investigating the impact of interactions
on the patterns of priority-queue dynamics. 
Oliveira and Vazquez introduced a model of two 
interacting priority queues with AND-type interaction \cite{ov}, 
and found that the waiting time distribution $P(\tau)$
of the model is still a power law for large $\tau$, 
yet with different exponents such as $\alpha_I=2$ for the $I$-tasks in the $L=2$ case.
Their model has been further generalized into the priority-queue 
{\it networks} of $N>2$ interacting queues \cite{min},
where it is found that the AND-type interaction is not
suitable for the network with loops when $N>2$, leading to
frozen, trivial queue dynamics.
Therefore a scalable interaction of OR-type rule is introduced 
in the priority-queue network (the OR model \cite{min}). The OR model
leads to different $P(\tau)$ than the AND-type one
and the power-law exponent $\alpha$ is found to
depend on various factors such as the network size, global network
topology, and local position of a queue node in a diverse way.
These works have demonstrated that the interaction is indeed a relevant and 
consequential factor in the priority-queue dynamics.

In this paper, we extend the work of Ref.~\cite{min},
and consider further forms of human interaction
in the priority-queue networks.
Here we are specifically interested in two factors: the team-based
task execution and the hierarchy in decision making.
The former refers to the situation when a task demands simultaneous actions
of more than two individuals and an individual's decision
is affected by more than one others.
The latter applies when there is a hierarchy in the queue nodes' status
in a way that a node in higher hierarchy can order the execution of
a task to the node in lower hierarchy. 
Both forms of human interaction are encountered in many real-life
situations, thus the understanding of their impact is essential 
for a more complete human dynamics modeling, as well as for
a more thorough understanding of the priority-queue network dynamics in general.

\begin{figure}
\vskip -5mm
\centerline{\epsfxsize=\linewidth \epsfbox{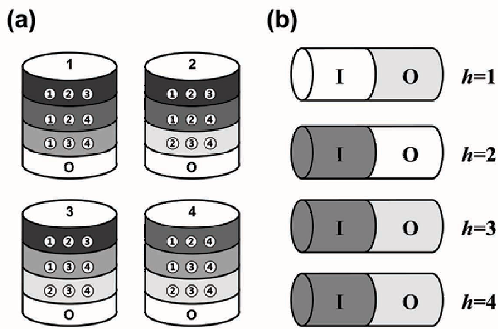}}
\vskip -5mm
\caption{(Color Online) Schematic illustration of the interaction
rules of the models.
{\bf (a)}
$(N,r)$-model for $N=4$ queues with teams of size $r=3$.
Each queue has ${3\choose2}$ $I$-tasks (shaded) and one $O$-tasks (white). 
Each team is indicated by different shade level and labeled with
team members.
{\bf (b)}
$H$-model for $N=4$. 
Each queue has one $I$-task and one $O$-task.
Arranged by the hierarchy value from top to bottom,
when the a node, say the $h=2$ node, executes its $I$-task, 
the nodes in the lower hierarchy, $h=3$ and $h=4$ nodes,
has to follow to execute their $I$-task (dark).
}
\end{figure}

\section*{Models}

The models studied in this paper are
built upon the priority-queue network model of Ref.~\cite{min},
with newly-introduced interaction rules.
Depending on the interaction rule, each queue node has
$\Theta_i$ $I$-tasks and an $O$-task $(L_i=\Theta_i+1)$.
Initially each task is given a priority value drawn from a uniform
distribution in $[0,1)$.
Then each step, a queue node is chosen randomly (say it to be $i$)
and its highest priority task is identified.
If it is an $I$-task, its execution is challenged against the queue discipline
(interaction rules) of the particular model.
If the selected $I$-task does pass the challenge,
the node $i$ and all other nodes involved in the task 
executes it simultaneously.
Otherwise, only the node $i$ executes its $O$-task instead.
Upon execution, the waiting time $\tau$ of the tasks is recorded, 
and the executed tasks are replaced with new
tasks each with a random priority value in uniform $[0,1)$.
$N$ such updates constitute a Monte Carlo step,
which is the time unit of waiting time measurement.

{\em Interaction rule in the team-based task execution---}
In the team-based task execution model, 
each $I$-task is associated with a group of queue nodes of size $r$, meaning that
more than two individuals are involved in the execution of the task.
Here the size of a team $r$ is the important parameter characterizing 
the model. We call it an $(N,r)$-model hereafter.
In the $(N,r)$-model, each queue $i$ has $\Theta_i={N-1\choose r-1}$ $I$-tasks
plus an $O$-task, thus is of equal fixed length of $L_i={N-1\choose r-1}+1$
(Fig.~1a). In this model, when a node (chosen randomly in each 
step) has the $I$-task, $I_G$, as its highest priority task,
all the nodes in the team $G$ executes it at the step; an OR-type rule.
The case with $r=2$ is the same as the original OR model with
the pairwise task execution of Ref.~\cite{min}. 
As $r>2$, the number of $I$-tasks $\Theta_i$ changes drastically,
which is expected to affect the queue network's dynamics
and thus is of interest in this work.
This model can be thought of as the generalization
of the priority-queue network model into the priority-queue hypergraphs.

{\em Interaction rule in the hierarchical decision---}
In the hierarchy-based task execution model (the $H$-model, hereafter),
$N$ individuals are assigned its hierarchy value from $h=1$ to $h=N$.
Each node has two tasks, one $I$-task and an $O$-task; thus $L_i=2$ (Fig.~1b).
When a node $i$ ($i$-th in the hierarchy) chooses its $I$-task
to be the highest priority, all the lower hierarchy nodes (from $i+1$ to $N$)
follow to execute it simultaneously with the node $i$.
Depending on the node's position in the hierarchy, the degree
of interruption (number of $I$-task calls from other nodes) varies, 
the effect of which is of interest in this work.

\section*{Results}
{\em Team-based task execution model: The $(N,r)$-model---}
In Fig.~2, we show the $P(\tau)$ of the $(N,r)$-model with $N=4,5,6$.
As mentioned above, $r=2$ case of the $(N,r)$-model is identical
to the OR model of Ref.~\cite{min} on the fully-connected network
and we obtain that $\alpha_{I}\approx2$ and $\alpha_O\approx 1.3$ (Fig.~2, {\small$\Box$}).
For $3\le r\le N-1$, the waiting time dynamics is found to have the same 
asymptotics as the $r=2$ case (Fig.~2, {$\triangledown$},~{\large$\diamond$},~{\Large$\circ$}).
Notable exception is the $I$-task dynamics for $r=N-1$, for which
the waiting time exponent is larger than the other cases as
$\alpha_{I,r=N-1}\approx2.5$ (Figs.~2a-c, {\Large$\circ$}).
Interestingly, this distinct power-law exponent for $r=N-1$ has also been observed
in the pairwise $(r=2)$ OR model on fully-connected networks. There
the case with $N=3$ exhibits a distinct power-law exponent than
other cases with $N\ge4$ \cite{min}. The origin of this distinct behavior of $r=N-1$
case is not fully understood yet, calling for a further study.
For the $O$-task dynamics, although the probability density for large
$\tau$ becomes elevated with $r$, the power-law exponent remains unchanged (Figs.~2d-f).

\begin{figure}
\begin{minipage}{0.5\linewidth}
\centerline{\epsfxsize=\linewidth \epsfbox{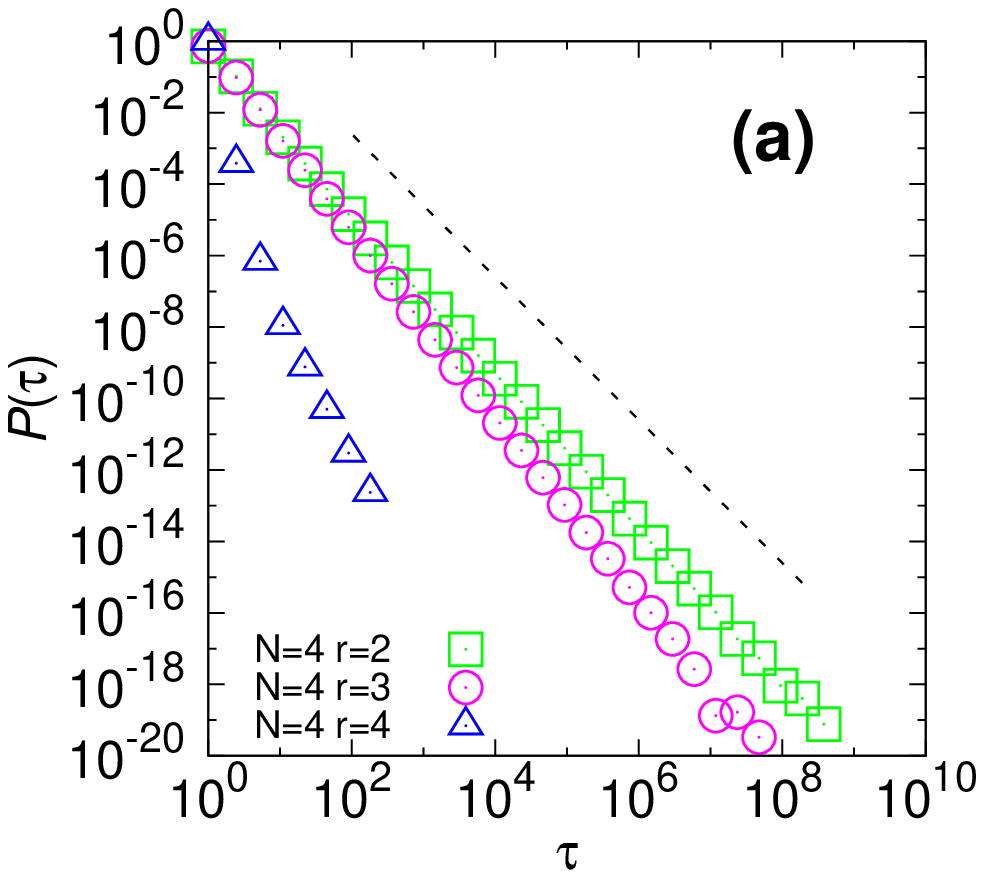}}
\centerline{\epsfxsize=\linewidth \epsfbox{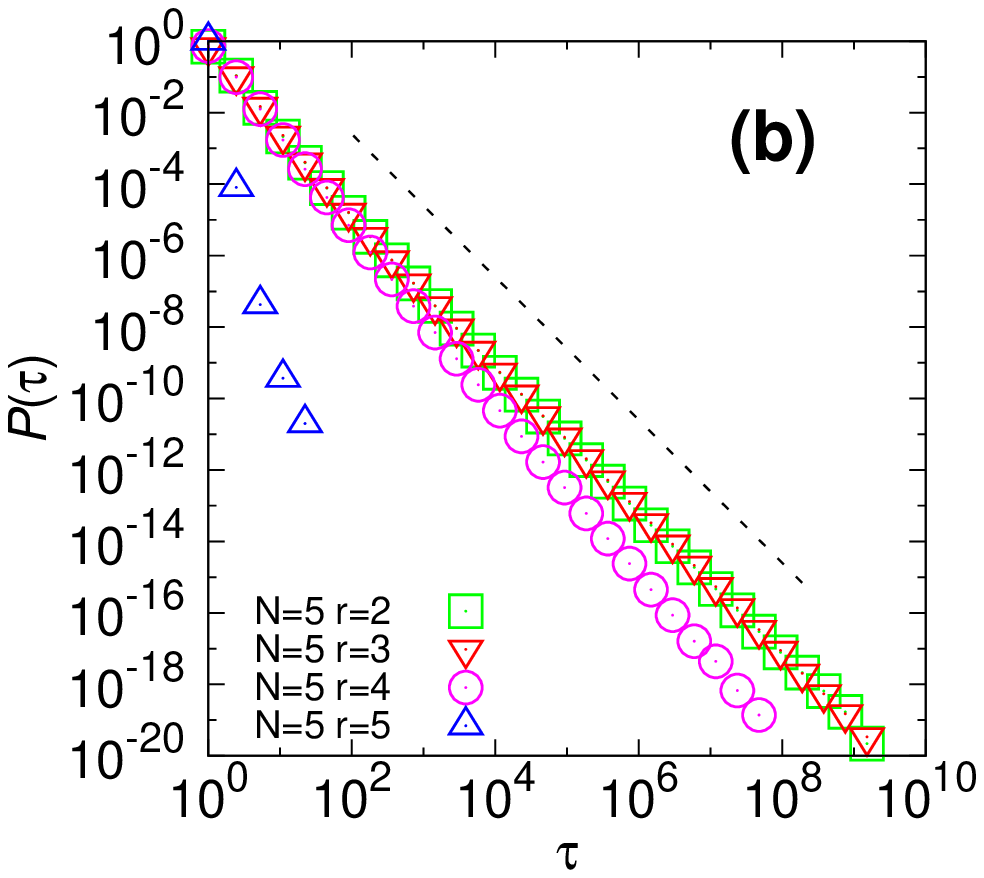}}
\centerline{\epsfxsize=\linewidth \epsfbox{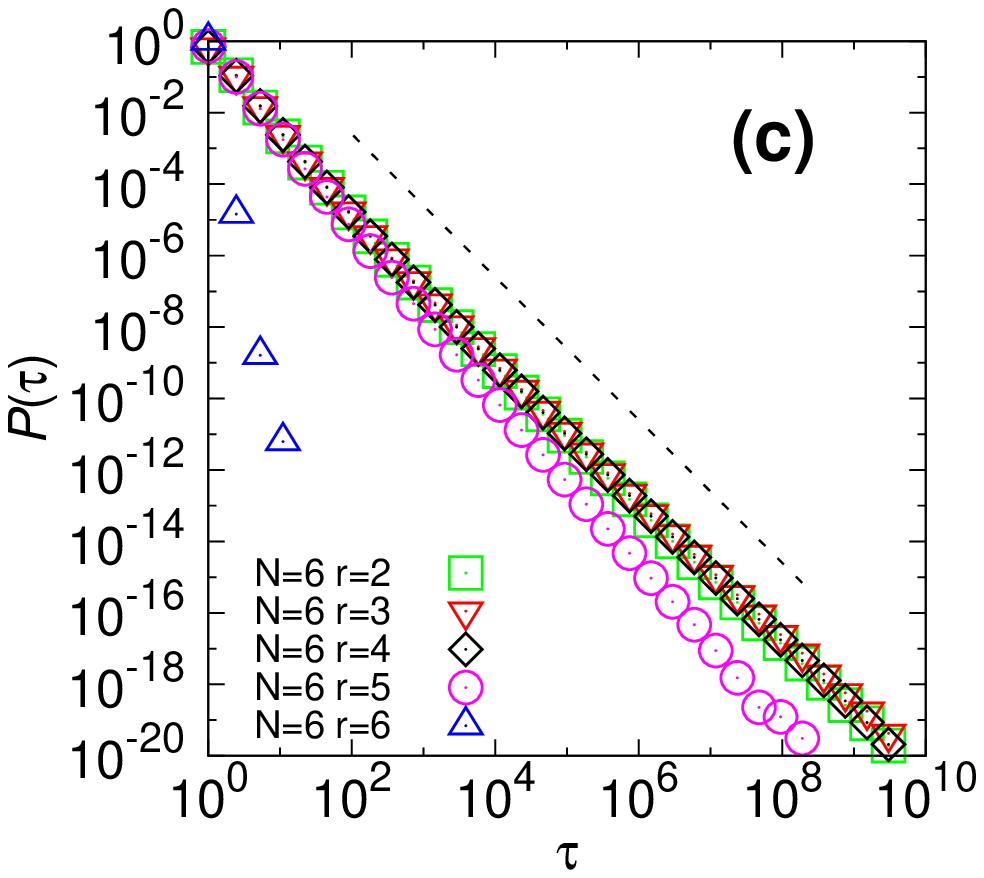}}
\end{minipage}\hfill
\begin{minipage}{0.5\linewidth}
\centerline{\epsfxsize=\linewidth \epsfbox{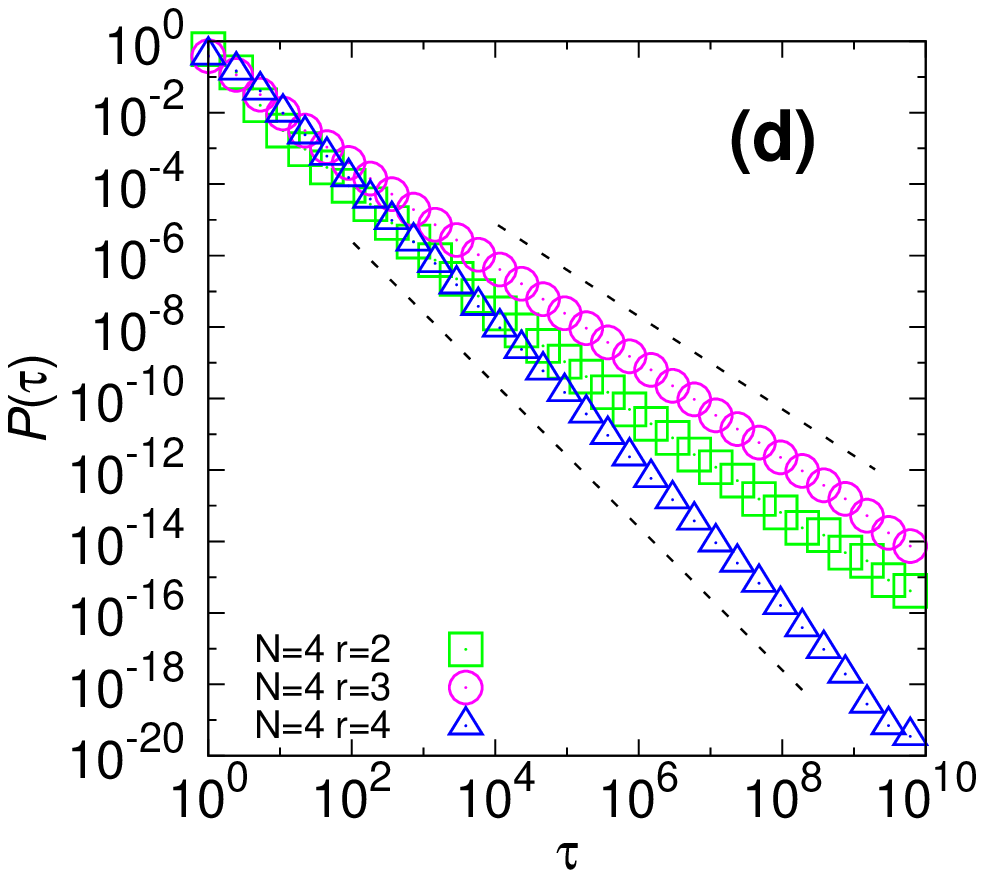}}
\centerline{\epsfxsize=\linewidth \epsfbox{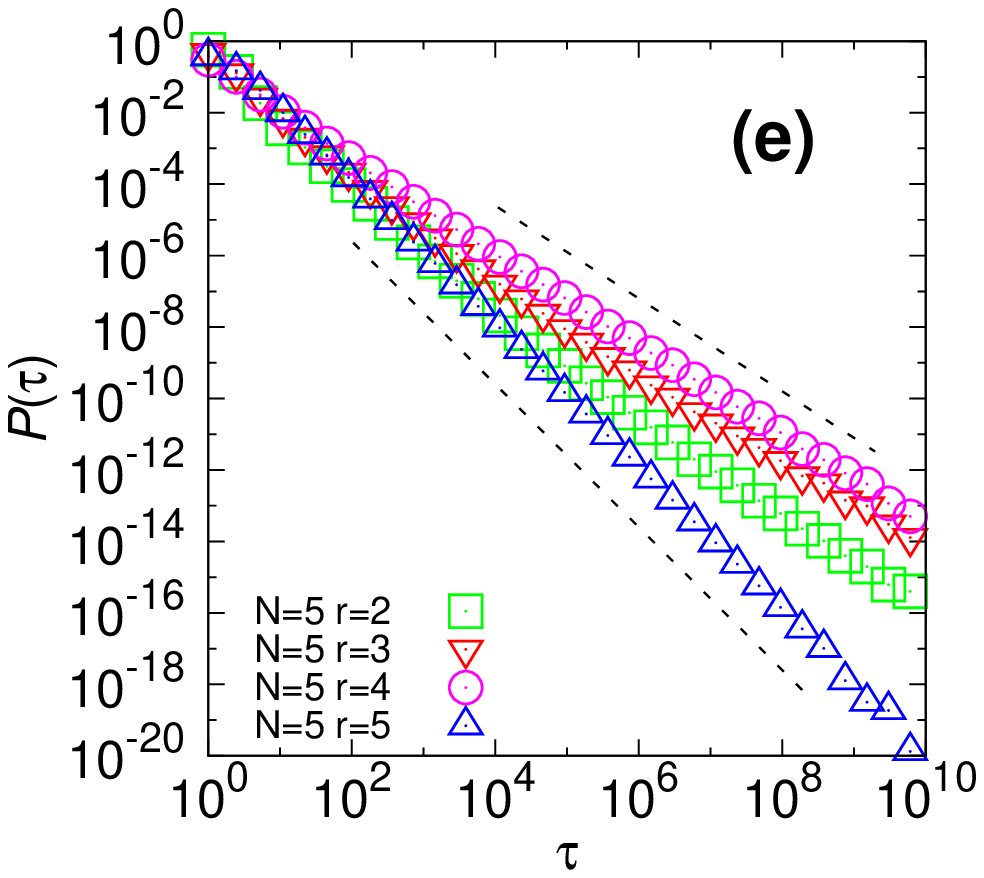}}
\centerline{\epsfxsize=\linewidth \epsfbox{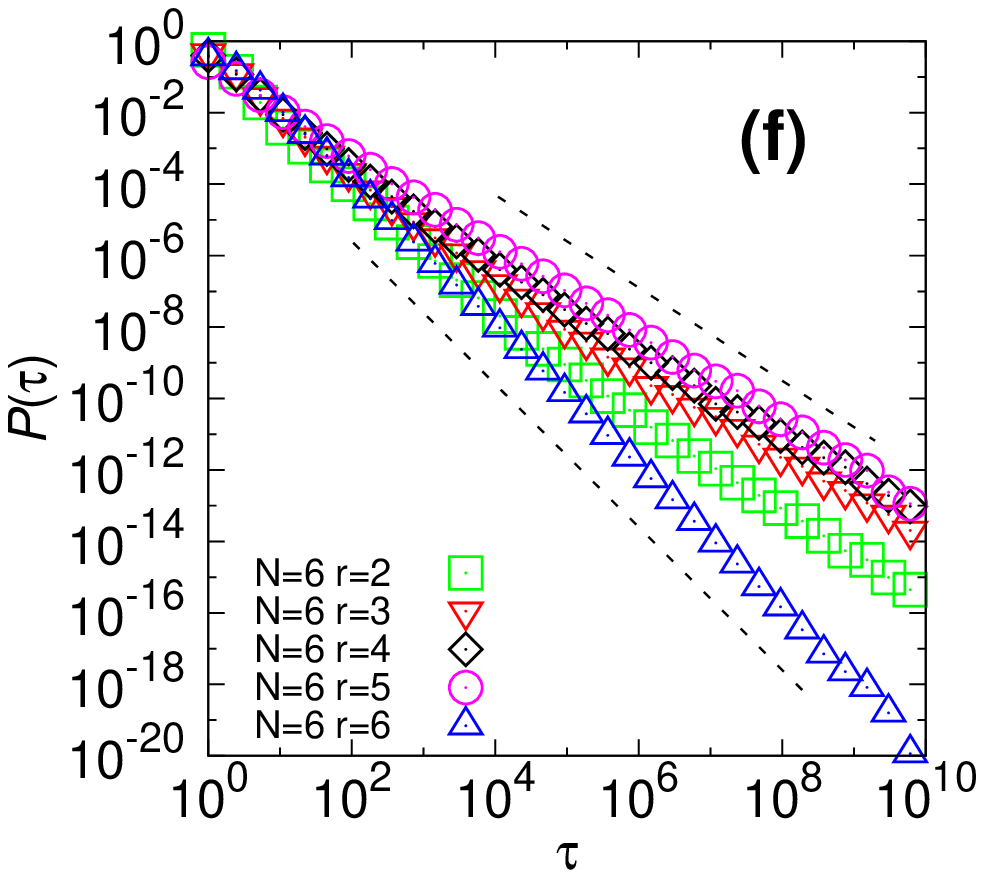}}
\end{minipage}
\caption{(Color Online) Waiting time distribution $P(\tau)$
of the team-based task execution model, the $(N,r)$-model.
{\bf (a--c)}
The waiting time distribution $P(\tau)$ of $I$-tasks for
various team size $r$ with $N=4$ (a), $5$ (b), and $6$ (c).
Slope of the straight lines is $-2$, drawn for the eye.
{\bf (d--f)}
The waiting time distribution $P(\tau)$ of $O$-tasks for
various team size $r$ with $N=4$ (d), $5$ (e), and $6$ (f).
Slopes of the straight lines are $-1.3$ (upper) and $-2$ (lower), 
respectively, drawn for the eye.
}
\end{figure}

The case with $r=N$ exhibits a distinct behavior.
Its $I$-task dynamics exhibits a fast decaying, exponential-like $P(\tau)$ 
and the $O$-task dynamics shows a power-law with the exponent 
$\alpha_{O,r=N}\approx2$, distinct from the cases with $r< N$ (Fig.~2, {\small$\triangle$}).
For $r=N$, each queue node has one $I$-task and one $O$-task, similar to
the $N=2$ OR model. However, the $I$-task should be executed if
any one queue has the $I$-task as its highest priority task,
giving a disproportionately high probability to $I$-task executions,
thereby leading to the exponential-like $P(\tau)$ for the $I$-task.
Meanwhile, from the perspective of the $O$-task, the situation is
not different from that of the $N=2$ OR model, thereby leading to
the same exponent $\alpha_O\approx2$ as that of the $N=2$ OR model \cite{min}.

In the $(N,r)$-model, the mean $I$-task waiting time is finite
but the variance diverges for $r<N$ $(\alpha_I<3)$.
For the $O$-task, not only the variance but also
the mean waiting time diverge $(\alpha\lesssim 2)$,
implying the lack of characteristic scale in the waiting times dynamics \cite{hklee}.

\begin{figure}
\begin{minipage}{0.5\linewidth}
\centerline{\epsfxsize=\linewidth \epsfbox{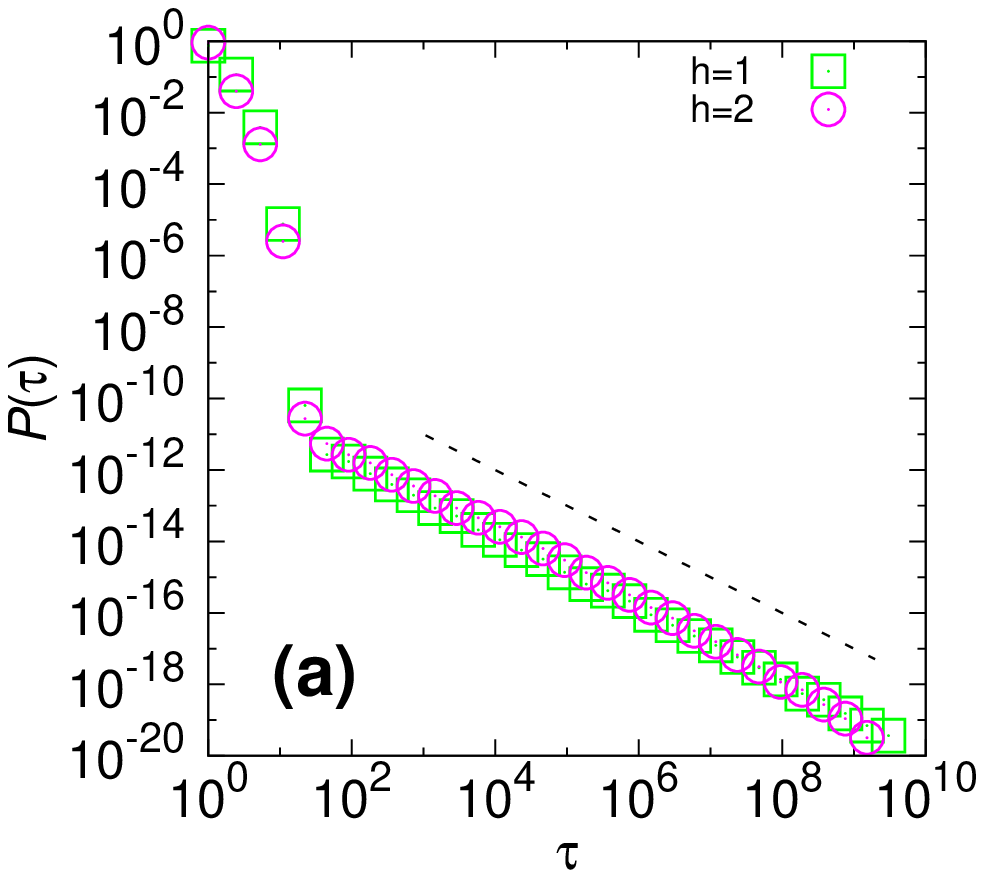}}
\centerline{\epsfxsize=\linewidth \epsfbox{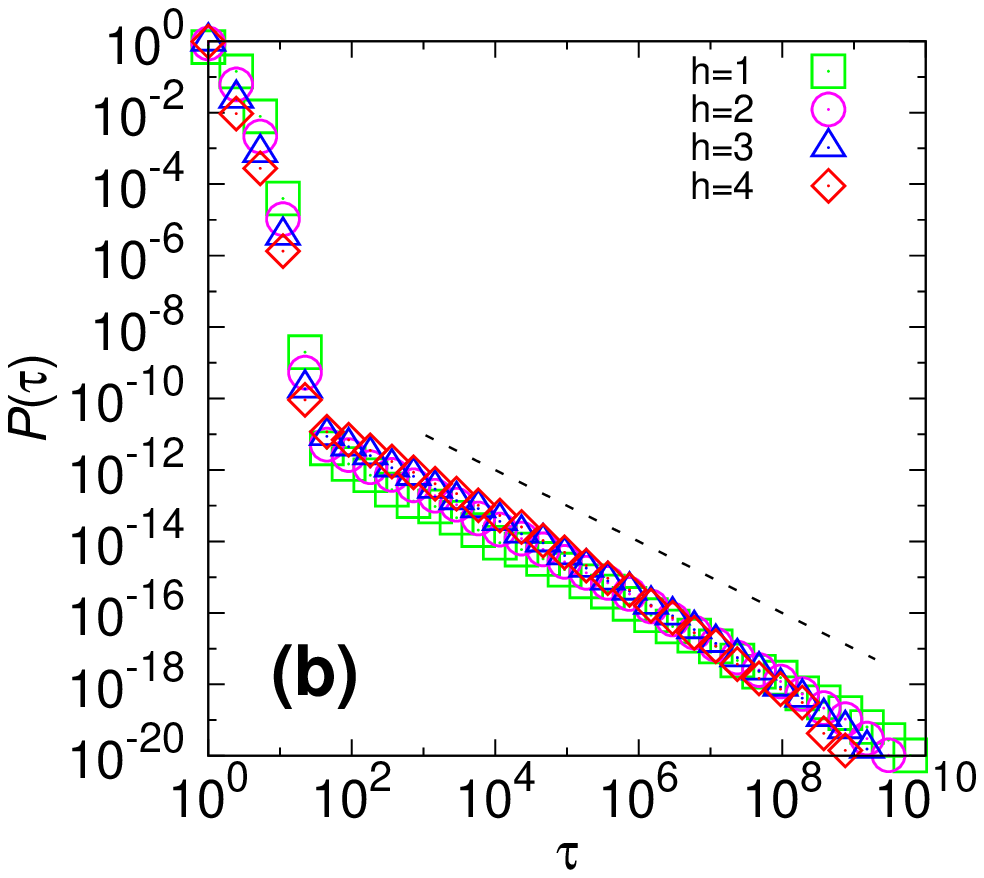}}
\centerline{\epsfxsize=\linewidth \epsfbox{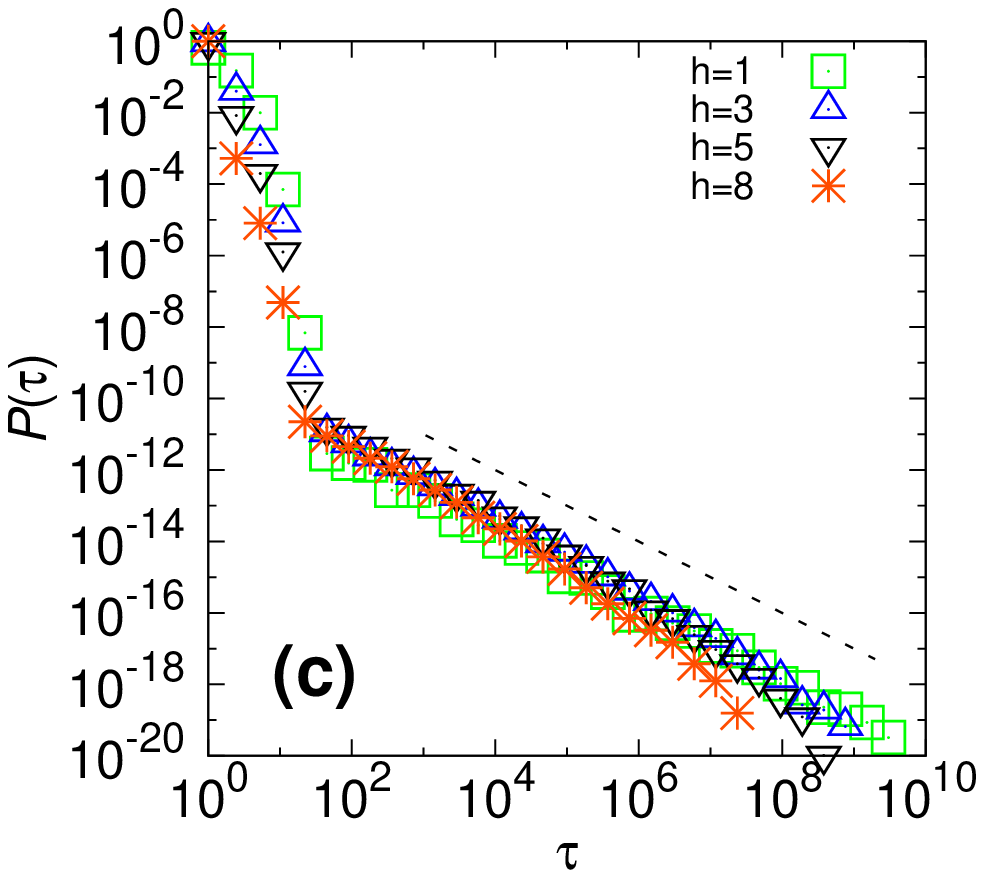}}
\end{minipage}\hfill
\begin{minipage}{0.5\linewidth}
\centerline{\epsfxsize=\linewidth \epsfbox{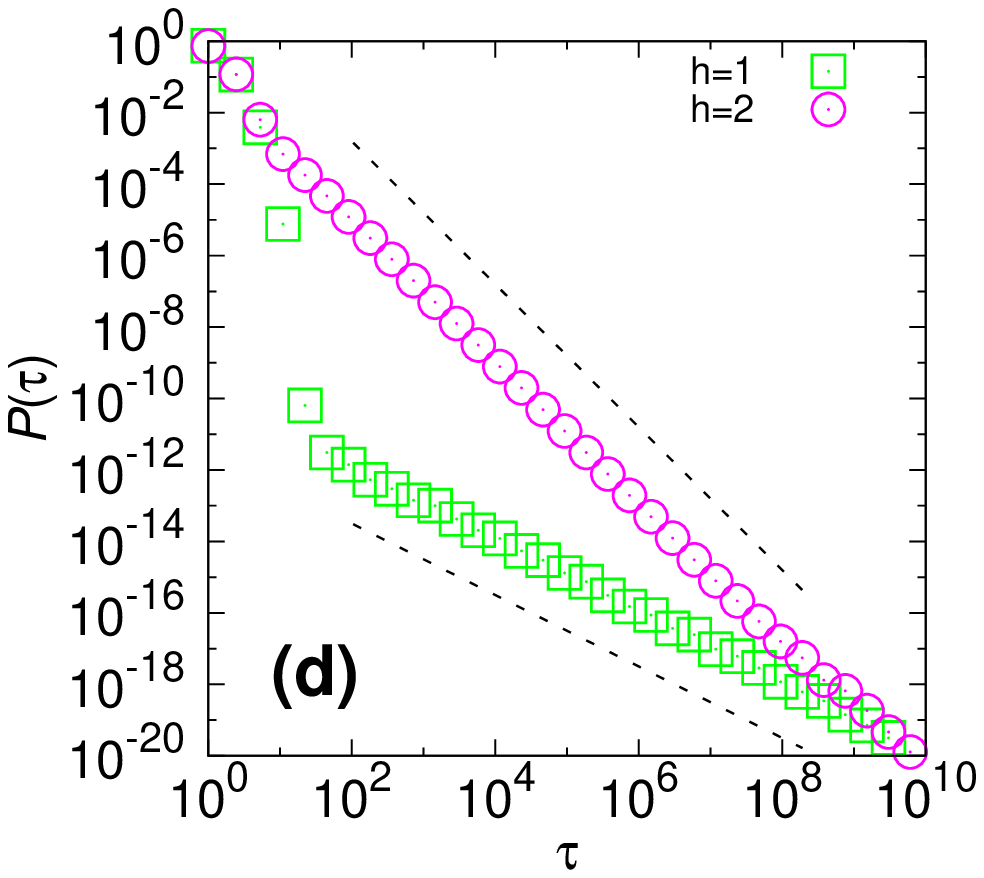}}
\centerline{\epsfxsize=\linewidth \epsfbox{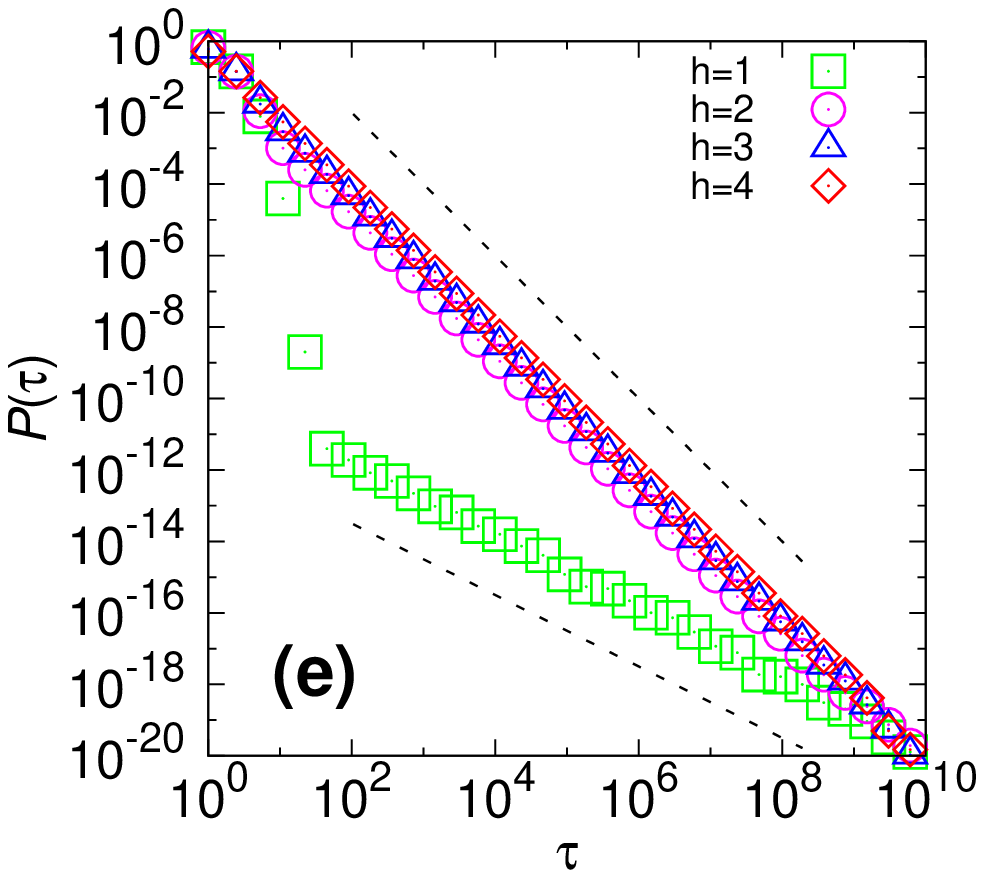}}
\centerline{\epsfxsize=\linewidth \epsfbox{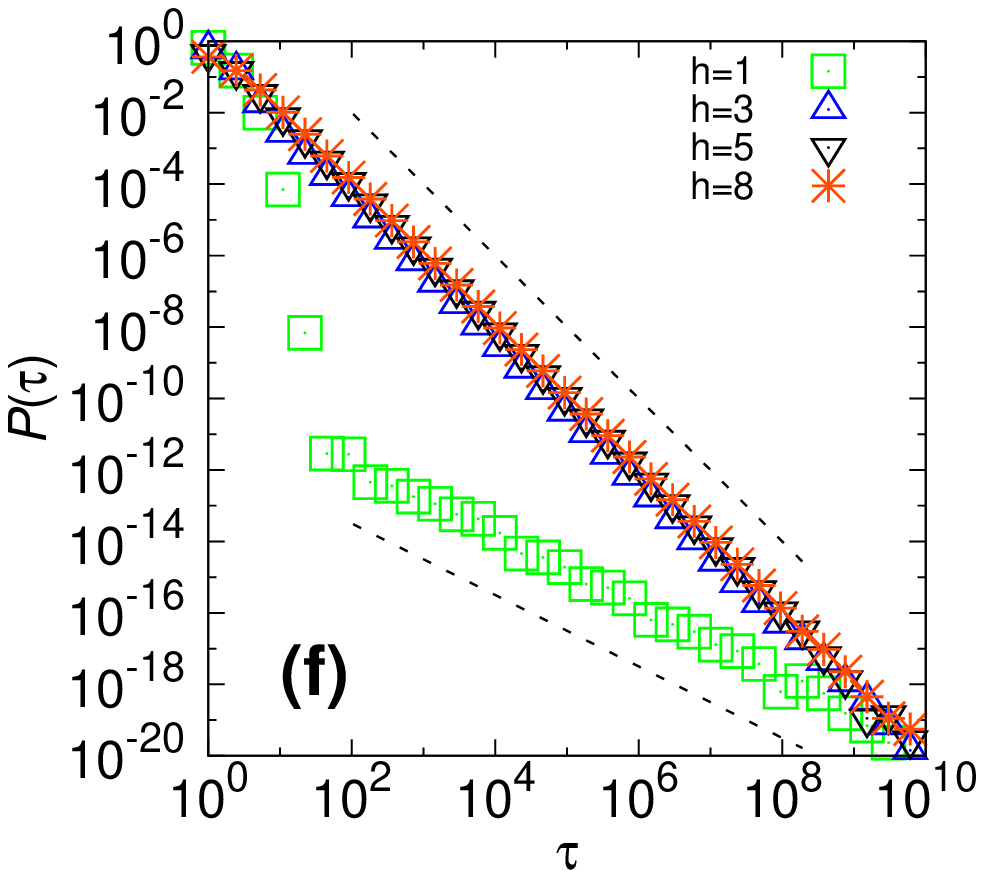}}
\end{minipage}
\caption{(Color Online) Waiting time distribution $P(\tau)$
of the hierarchical decision model, the $H$-model.
{\bf (a--c)}
The waiting time distribution $P(\tau)$ of $I$-tasks for
various hierarchy value $h$ with $N=2$ (a), $4$ (b), and $8$ (c).
Slope of the straight lines is $-1$, drawn for the eye.
{\bf (d--f)}
The waiting time distribution $P(\tau)$ of $O$-tasks for
various hierarchy value $h$ with $N=4$ (d), $5$ (e), and $6$ (f).
Slopes of the straight lines are $-2$ (upper) and $-1$ (lower), 
respectively, drawn for the eye.
}
\end{figure}

{\em Hierarchical decision model: The $H$-model---}
In Fig.~3, we show the $P(\tau)$ of the $H$-model with
$N=2,4,$ and $8$.
In the $H$-model, the node at the top of hierarchy ($h=1$)
is not affected by any other nodes in its decision making.
Thus the $h=1$ node's dynamics, having one $I$-task and one $O$-task,
is the same as that of the Barab\'asi queue with length $L=2$ \cite{alb}.
Thus we have $\alpha=1$ for both the $I$-task and $O$-task
for the $h=1$ node (Fig.~3, {\small$\Box$}).

The $I$-task dynamics of nodes with $h>1$ is found to follow
the power-law $P(\tau)$ with the same exponent $\alpha_I\approx1$,
although these nodes' $I$-task dynamics is affected
by the nodes in the higher hierarchy. This behavior can 
be understood by the following argument.
In the $H$-model, a node's $I$-task dynamics is the superposition
of subdynamics having power-law $P(\tau)$ with tail exponent $-1$,
originating from the node itself and its higher hierarchy nodes.
Due to the singular nature of the power-law distribution with
the exponent $-1$, the superposed dynamics
maintains the same tail exponent $\alpha\approx1$,
upto the characteristic $\tau$ which decreases with $h$, namely 
the number of superposed subdynamics.
Therefore,
$I$-task dynamics of the $H$-model follows the power-law $P(\tau)$
with the same exponent $\alpha_I\approx1$ independent of
node's position in the hierarchy. Meanwhile, although
the power-law exponent $\alpha_I$ is close to $1$,
the mean waiting time is finite, as the mean value is dominated by 
the strong peak of $P(\tau)$ at $\tau=1$. However, the power-law
tail with $\alpha_I\approx1$ means that once the $I$-task
is shelved incidentally, it may wait for an extremely long time
before execution.

The $O$-task dynamics of nodes with $h>1$ is
affected more significantly by the hierarchical decision.
For these nodes, the relative priority of the $O$-task
can be changed by the forced execution of the $I$-task
following the call from the higher hierarchy nodes.
Thus the situation becomes similar to the $O$-task dynamics
in the OR model with $N=2$ nodes having $\alpha_O\approx2$, 
where the $I$-task priority can be updated via the $I$-task call 
from the other node regardless of its own priority value.
This similarity explains the exponent $\alpha_O\approx2$
for the $h>1$ nodes in the $H$-model. 
Whereas for the highest hierarchy $(h=1)$ node the mean $O$-task waiting time
is finite, for $h>1$ nodes it is marginally diverging $(\alpha_O\approx2)$, 
suggesting a lack of the characteristic waiting time.

\section*{Summary}
In this paper, we have considered two additional forms of interaction
in the priority-queue network models, motivated by
the team-based task execution and hierarchy-based decision-making,
which are encountered in many real-life human activities.
Our numerical study has shown 
that these generalizations maintain the power-law decaying waiting
time distribution $P(\tau)$, Eq.~(1), yet yield different values 
of the exponent $\alpha$ depending on the model parameters.

Most of the power-law behaviors exhibited by the team-based $(N,r)$-model 
are found to have a counterpart in the pairwise OR model's dynamics.
This suggests that the two-body interactions 
may contain the most fundamental dynamic consequence of interactions 
in the priority-queue network dynamics, although the precise mechanism of 
these correspondences has yet to be understood, due to the lack of analytic results.
For the hierarchical decision-based $H$-model,
it is found somewhat counterintuitively that 
it is the $O$-task dynamics, rather than the $I$-task dynamics,
that is more strongly affected due to the hierarchical, directional interaction,
compared to the reciprocal interaction case of the pairwise OR model
in which both the $I$- and $O$-task dynamics are affected.
It thus highlights the nontrivial role 
of reciprocity of interaction in the priority-queue dynamics.

\begin{acknowledgments}
This work was supported by the Korea Research Foundation Grant funded by 
the Korean Government (MEST) (KRF-2007-331-C00111).
\end{acknowledgments}

\end{document}